\documentclass[aps,prd,preprint,showpacs,showkeys,superscriptaddress]{revtex4}

\usepackage{latexsym}
\usepackage[dvips]{graphicx}

\begin{document}


\title{Entropy of the FRW cosmology based on the brick wall method}

\author{Wontae Kim}
\email[]{wtkim@sogang.ac.kr}
\affiliation{Department of Physics, Sogang University, Seoul 121-742, Korea}
\affiliation{Basic Science Research Institute, Sogang University, Seoul 121-742, Korea}
\affiliation{Center for Quantum Spacetime, Sogang University, Seoul 121-742, Korea}

\author{Edwin J. Son}
\email[]{eddy@sogang.ac.kr}
\affiliation{Department of Physics, Sogang University, Seoul 121-742, Korea}
\affiliation{Basic Science Research Institute, Sogang University, Seoul 121-742, Korea}

\author{Myungseok Yoon}
\email[]{younms@sogang.ac.kr}
\affiliation{Center for Quantum Spacetime, Sogang University, Seoul 121-742, Korea}

\date{\today}

\begin{abstract}
The brick wall method in calculations of the entropy of black holes 
can be applied to the FRW cosmology in order to
study the statistical entropy. An appropriate cutoff satisfying
the covariant entropy bound can be
chosen so that the entropy has a definite bound.
Among the entropy for each of cosmological eras,
the vacuum energy-dominated era turns out to give
the maximal entropy which is in fact compatible with assumptions from
the brick wall method.  
\end{abstract}



\maketitle


The holographic principle in quantum gravity suggested by 't Hooft
\cite{thooft:holography} shows that degrees of freedom of a
spatial region reside on its boundary. In the thermodynamics of black
holes, it has been known that the
entropy of a black hole is proportional to the area of the horizon
\cite{bekenstein,hawking}.  Recently, the holographic principle has
been studied in the cosmological context with a particle horizon
\cite{fs}. In the cosmology, the choice of holographic boundary is
in some sense unclear; in contrast to the black hole physics, for
which there is a definite boundary called event
horizon. However, it has been shown that the
thermodynamic first law $dU = TdS + W dV$ can be satisfied if the boundary
is chosen to be the cosmological apparent horizon, where $U$, $S$, $V$,
$T$, and $W$ are the internal energy, the entropy, the volume, the
temperature, and the work density, respectively
\cite{br,ck,swc,zrm}. This result has been induced from 
Einstein field equations when the temperature is defined by $T =
|\kappa|/2\pi$ with the entropy $S = {\cal A}/4G$, where $\kappa$,
$G$, and $\cal A$ are the surface gravity, the
gravitational constant, and the area of the apparent horizon,
respectively.

On the other hand, there are various methods to obtain the entropy
proportional to the area of the event horizon of a black hole. 
One of the most convenient way to get the statistical entropy 
is to use the brick-wall method
\cite{thooft}, where the cutoff parameter is introduced because
of divergence near the event horizon, and it has been applied to various
black holes \cite{mukohyama,su,dlm,gm,kkps,hkps}. Since
degrees of freedom of a field are dominant near horizon, the
brick-wall model has often been replaced by a thin-layer model, which
makes the calculation of entropy simple \cite{hzk,zl}.
Recently, it has been shown that the (thin-layered) brick-wall model can
be applied to a time-dependent black hole with an assumption of local
equilibrium near horizon~\cite{xz}.

Now, we would like to consider the entropy of
Friedmann-Robertson-Walker (FRW) universe based on 
the brick-wall model, since there exists the
apparent horizon satisfying the thermodynamic first law.
Moreover, this is the first application of this method to this
cosmological model as far as we know. 
For this purpose, the
apparent horizon of the universe will be introduced as the holographic boundary
instead of the event horizons of black holes, and the entropy of the FRW
universe will be formulated by means of the brick-wall method without 
specific solutions of the scale factor without loss of generality. 
Then, the entropy is calculated for the
various eras of the universe, and it is explicitly shown that the entropy can
never exceed the gravitational entropy with the same boundary,
$S\le S_g=\mathcal{A}/4G$, which is supported by the covariant
entropy bound(CEB)~\cite{bousso}.


Let us now start with the standard FRW metric, 
\begin{equation}
ds^2 = - dt^2 + a^2(t) \left[ \frac{dr^2}{1-kr^2}+r^2 d\Omega^2 \right], \label{FRW}
\end{equation}
where $k=0,\pm1$ are normalized spatial curvatures, and
$d\Omega^2=d\theta^2+\sin^2\theta d\phi^2$ is the line element of unit
two-sphere. To get the statistical entropy by the brick wall
method~\cite{thooft}, 
we change the radial
coordinate to $R=ra$ for convenience, then the
metric~(\ref{FRW}) can be written in the form of
\begin{equation}
ds^2 = - \left( \frac{1-R^2/R_A^2}{1-kR^2/a^2} \right) dt^2 -
\frac{2HR}{1-kR^2/a^2} dtdR + \frac{dR^2}{1-kR^2/a^2} + R^2 d\Omega^2,
\label{met}
\end{equation}
where $H=\dot{a}/a$ is the Hubble parameter, and the apparent horizon
$R_A$ is given by
\begin{equation}
R_A = \frac{1}{\sqrt{H^2+k/a^2}}. \label{AH}
\end{equation}
Then, the Klein-Gordon equation $[ \Box - \mu^2 ] \Psi=0$ in this
background is explicitly
written as
\begin{equation}
\left[ \partial_t \frac{-\partial_t-HR\partial_R}{\sqrt{1-kR^2/a^2}} -
  \frac{1}{R^2}\partial_R
  \frac{HR\partial_t-(1-R^2/R_A^2)\partial_R}{R^{-2}\sqrt{1-kR^2/a^2}}
  - \frac{\mu^2+R^{-2}\ell(\ell+1)}{\sqrt{1-kR^2/a^2}} \right]
\psi(t,R) = 0,
\end{equation}
and separation of variables is possible up to $\Psi=\psi(t,R)Y_{\ell m}(\theta,\phi)$, 
whereas $\psi(t,R)$ is no longer separable.
Note that the FRW cosmology also obeys the holographic
principle~\cite{fs,br,ck}, and it is sufficient to
investigate fields in the vicinity of horizon in order to get
the statistical entropy. Now, we assume the frequency to be constant
near horizon as like time-dependent Vaidya metric 
in Ref.~\cite{xz}, so that the field becomes
$\psi(t,R) = e^{-i\omega t + i v(t,R)}$.
Using the WKB approximation, $R\sim R_A\gg1$ along with slowly varying condition,
$\dot{v}\ll\omega$, the radial momentum can be defined by
\begin{equation}
v' \simeq \frac{-HR\omega \pm
  \sqrt{(1-kR^2/a^2)\omega^2-(1-R^2/R_A^2)[\mu^2+R^{-2}\ell(\ell+1)]}}{1-R^2/R_A^2} \equiv k_R^\pm, \label{mom}
\end{equation}
where the plus(minus) sign represents the out(in)-going wave. 
Note that the phase velocity of wave is $v_p=\omega/k_R^\pm$ in this approximation.

Actually, the present nonstatic cosmological model should be regarded as a nonequilibrium
system. However, the notion of local equilibrium 
as a working hypothesis may be used in the nonequilibrium
thermodynamics~\cite{kreuzer}. 
So, we will assume that the temperature of thermal radiations is slowly
varying near the apparent horizon. In fact, the 
temperature at the apparent horizon is approximately proportional 
to the inverse of the apparent horizon, $T\sim R_A^{-1}$, 
which will be explicitly calculated in later, and then 
the local equilibrium requires that
\begin{equation}
\frac{\delta T}{T}\sim\frac{\delta R_A}{R_A}\sim\frac{\delta a}{a}\ll1,
\end{equation}
where $\delta$ represents fluctuation of each quantity, and the
inequality is satisfied as far as the Hubble parameter is very small,
$H\ll1$.

Next, the number of radial modes according to the semiclassical
quantization rule is given by
\begin{eqnarray}
2\pi n &=& \int_{R_A-h-\delta}^{R_A-h} dR\ k_R^+ +
  \int_{R_A-h}^{R_A-h-\delta} dR\ k_R^- \nonumber \\
&=& 2 \int_{R_A-h-\delta}^{R_A-h}
  dR\frac{\sqrt{(1-kR^2/a^2)\omega^2-(1-R^2/R_A^2)[\mu^2+R^{-2}\ell(\ell+1)]}}{1-R^2/R_A^2},
\end{eqnarray}
where $h$ and $\delta$ are cutoffs, and both of them are assumed to be
very small positive quantity compared to the apparent horizon,
$h,\delta\ll R_A$. 
Then, the total
number of modes for given energy $\omega$ is given by
\begin{equation}
N = \int d\ell (2\ell+1) n,
\end{equation}
where the integration goes over those values for which the square root
in the radial momentum~(\ref{mom}) is real.
Following 't Hooft~\cite{thooft}, the free energy is given by
\begin{eqnarray}
\beta F 
&=& \int dN \ln \left( 1-e^{-\beta\omega}\right) \nonumber \\
&=& -\int d\omega \frac{\beta N}{e^{\beta\omega}-1} \nonumber \\
&=& -\frac{\beta}{\pi} \int_{R_A-h-\delta}^{R_A-h} dR
  \left(1-\frac{R^2}{R_A^2}\right)^{-1} \int d\omega
  \left(e^{\beta\omega}-1\right)^{-1} \int d\ell (2\ell+1) \nonumber \\
& & \qquad \times \sqrt{(1-kR^2/a^2)\omega^2-(1-R^2/R_A^2)[\mu^2+R^{-2}\ell(\ell+1)]}.
\end{eqnarray}
Now, in the approximation of
$\mu^2 \ll R_A/\beta^2\epsilon$, $\epsilon \ll 1/H^2R_A$,
where $\epsilon=h,h+\delta$, the free energy in the leading order is
simplified as
\begin{equation}
F \simeq -\frac{\pi^3}{90} (HR_A)^3 \left(\frac{R_A}{\beta}\right)^4 \frac{\delta}{h(h+\delta)},
\end{equation}
and then the main contribution of the apparent horizon to the internal
energy $U$ and the entropy $S$ are given by
\begin{eqnarray}
U &=& \frac{\partial}{\partial\beta} (\beta F) \simeq \frac{\pi^3}{30}
  (HR_A)^3 \left(\frac{R_A}{\beta}\right)^4 \frac{\delta}{h(h+\delta)},
  \label{energy} \\
S &=& \beta(U-F) \simeq \frac{2\pi^3}{45}
  \left(\frac{HR_A^2}{\beta}\right)^3 \frac{R_A\delta}{h(h+\delta)},
  \label{ent}
\end{eqnarray}
respectively, where we used the unified first law, $dU=TdS+WdV$.
Since proper lengths of cutoffs $h$
and $\delta$ are given by
$\bar{h} = \int_{R_A-h}^{R_A} dR/\sqrt{1-R^2/R_A^2} \simeq \sqrt{2R_Ah}$,~ 
$\bar{\delta} = \int_{R_A-h-\delta}^{R_A-h}
  dR/\sqrt{1-R^2/R_A^2} \simeq \sqrt{2R_A(h+\delta)} -
  \sqrt{2R_Ah}$,
the entropy~(\ref{ent}) reads
\begin{equation}
S \simeq \frac{\pi^2}{45} \left(\frac{HR_A^2}{\beta}\right)^3
  \frac{\mathcal{A}}{\bar{h}^2} \label{entropy}
\end{equation}
for an arbitrary $\bar\delta$ satisfying $\bar{h}\ll\bar\delta\ll
R_A$, where $\mathcal{A}=4\pi R_A^2$ is the area of the apparent
horizon which depends on time.

Next, we calculate thermodynamic quantities for some cosmological eras.
For this purpose, let us consider the equations of motion for the FRW cosmology,   
\begin{eqnarray}
& & H^2+\frac{k}{a^2} = \frac{8\pi G}{3}\rho + \frac13\Lambda, \\
& & \dot{H} - \frac{k}{a^2} = -4\pi G (\rho+p),
\end{eqnarray}
where $\rho$ is the energy density, $p$ is the pressure, and $\Lambda$
is the cosmological constant. They can be recast in the form of
\begin{eqnarray}
& & H^2 
= \frac{8\pi G}{3} (\rho_\mathrm{tot}+\rho_k), \label{eom} \\
& & \dot{H} + H^2 
= \frac{4\pi G}{3} (1+3\gamma) \rho_\mathrm{tot}, \label{con:en}
\end{eqnarray}
where $\rho_\mathrm{tot}=\rho+\rho_\Lambda=\rho_m+\rho_\mathrm{rad}+\rho_\Lambda$, $\rho_\Lambda=\Lambda/8\pi G$,
$\rho_k=-3k/8\pi Ga^2$.
The equation-of-state parameter $\gamma=p/\rho$ for each type of
energy is listed in table \ref{tab:energy} for convenience; in particular, $\gamma$ in
Eq.~(\ref{con:en}) is defined by
\begin{equation}
\gamma = \frac{p_\mathrm{tot}}{\rho_\mathrm{tot}}
= \frac{\Omega_\mathrm{rad}/3-\Omega_\Lambda}{\Omega_\mathrm{tot}}, \label{eos}
\end{equation}
where $p_\mathrm{tot}=p+p_\Lambda=p_m+p_\mathrm{rad}+p_\Lambda$.
The density parameter for a certain type of energy $\rho_i$
($i=m,\mathrm{rad},\Lambda$, etc.) was defined by $\Omega_i=\rho_i/\rho_c$,
and $\rho_c=3H^2/8\pi G=\rho_\mathrm{tot}+\rho_k$ is the critical density.
\begin{table}[tdp]
\begin{center}
\begin{tabular}{|c|c|c|c|c|c|}
\hline
& rad & $m$ & $k$ & $\Lambda$ & p-law \\
\hline
$\rho\sim$ & $a^{-4}$ & ~$a^{-3}$~ & $a^{-2}$ & $a^0$ & $a^{-n}$ \\ 
\hline
~$\gamma=$~ & ~1/3~ & 0 & ~$-1/3$~ & ~$-1$~ & ~$n/3-1$~ \\ 
\hline
\end{tabular}
\end{center}
\caption{
The energy density $\rho$ and equation-of-state parameter
$\gamma=p/\rho$ are calculated for radiation-, matter($m$)-, spatial
curvature($k$)-, and vacuum energy($\Lambda$)-dominated universes. In
general, for an energy density evolving as power-law, $\rho\sim
a^{-n}$, the equation-of-state parameter is given by
$\gamma=n/3-1$~\cite{carroll}.}
\label{tab:energy}
\end{table}
Here, the subscript `tot' means the sum of actual energy density
and pressure, excluding the contributions from the spatial
curvature~\cite{carroll}.


Now, the inverse of the temperature of a system on the apparent horizon is
defined by $T=\beta^{-1}=|\kappa|/2\pi$, where
$\kappa=(2\sqrt{-h})^{-1}\partial_a\sqrt{-h}h^{ab}\partial_bR|_{R=R_A}$
is the surface gravity at the apparent horizon, $h_{ab}$ is defined by
$ds^2=h_{ab}dx^adx^b+R^2(x)d\Omega^2$, and $a,b=0,1$~\cite{hayward}.
Then, the temperature for the FRW cosmology is calculated as
\begin{eqnarray}
\beta^{-1} &=& \frac{H^2R_A}{2\pi} \left| 1 + \frac12
  \left(\frac{\dot{H}}{H^2}+\frac{k}{H^2a^2}\right) \right| \nonumber \\ 
&=& \frac{1}{2\pi R_A} \left| \frac{1-3\gamma}{4} \right|, \label{T}
\end{eqnarray}
where we used the equations of motion~(\ref{eom}) and (\ref{con:en}).
The apparent horizon~(\ref{AH}) in Eq. (\ref{T}) was written in the
form of
\begin{equation}
R_A = \frac{1}{H\sqrt{1-\Omega_k}} = \frac{1}{H\sqrt{\Omega_\mathrm{tot}}}.
\end{equation}
From now on, we have to assume $\Omega_k<1$ so that the apparent horizon is
well-defined and the total energy density is positive,
$\Omega_\mathrm{tot}=1- \Omega_k>0$.
Then, the internal energy~(\ref{energy}) and the entropy~(\ref{entropy})
are rewritten as
\begin{eqnarray}
U &=& \frac{R_A}{240\pi\Omega_\mathrm{tot}^{3/2}\bar{h}^2}
  \left|\frac{1-3\gamma}{4}\right|^4 = \frac{3\mathcal{F}^{4/3}}{4}
  \mathcal{M}, \label{U} \\
S &=& \frac{\mathcal{A}}{360\pi\Omega_\mathrm{tot}^{3/2}\bar{h}^2}
  \left|\frac{1-3\gamma}{4}\right|^3
= \mathcal{F} \frac{\mathcal{A}}{4G}, \label{S}
\end{eqnarray}
when we set the cutoff as
\begin{equation}
\bar{h} = \frac{\ell_P}{\sqrt{90\pi}\Omega_\mathrm{tot}^{3/4}}, \label{cutoff}
\end{equation}
%
where $G=c^3\ell_P^2/\hbar$, $\ell_P$ is the Planck length, $\mathcal{M}=R_A/2G$ is the
Misner-Sharp energy~\cite{ms}, and the factor $\mathcal{F}$ is given by
\begin{equation}
\mathcal{F} = \left(\frac{|1-3\gamma|}{4}\right)^3. \label{F}
\end{equation}

\begin{figure}[pbt]
  \includegraphics[width=0.4\textwidth]{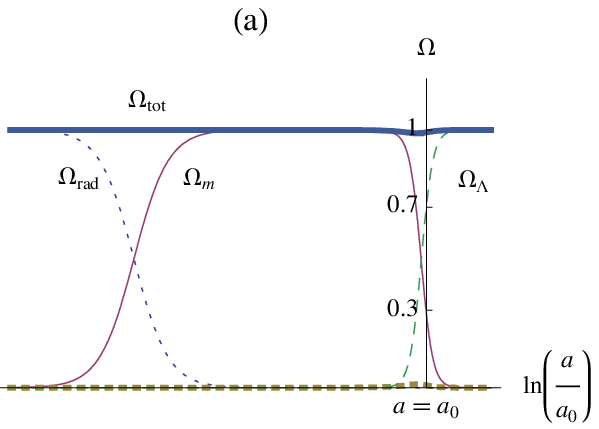}
  \hspace{0.05\textwidth}
  \includegraphics[width=0.4\textwidth]{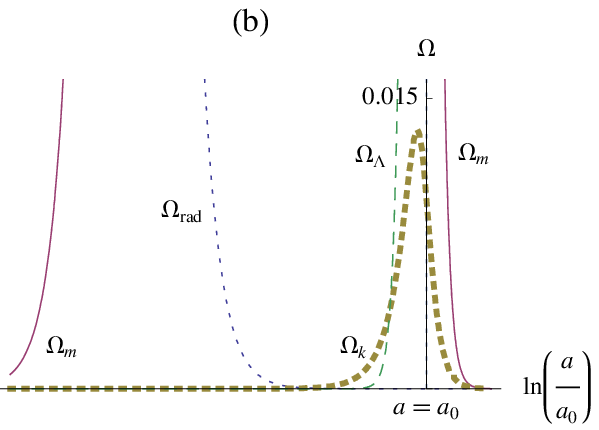}
  \caption{\label{fig:DP}The density parameters
  $\Omega_\mathrm{tot}$(thick, solid line),
  $\Omega_\mathrm{rad}$(thin, dotted line), $\Omega_m$(thin, solid
  line), $\Omega_\Lambda$(dashed line), and $\Omega_k$(thick, dotted
  line) are shown with respect to the logarithmic scale, where $a_0$
  represents the present scale of the universe. The profiles are
  plotted for $k=-1$, but they are not so much different from the case
  of $k=1$; simply speaking, $\Omega_k$ changes its sign and
  $\Omega_\mathrm{tot}$ becomes slightly large than 1 near the present
  scale $a_0$. (a) It shows that $\Omega_\mathrm{tot}\simeq1$. 
(b) In order to see the detail profile of $\Omega_k$, we magnify the bottom of (a) with the same scale in the horizontal axis.}
\end{figure}

Note that the cutoff~(\ref{cutoff}) may depend on time, since the density parameter
$\Omega_\mathrm{tot}=1-\Omega_k=1+k/H^2a^2$ can be time-dependent, 
for instance, for the case of $k=\pm1$.
In fact, it is due to the nonstatic apparent horizon as
discussed in Ref.~\cite{xz}. However, since the recent observational data~\cite{wmap5}
leads to the density parameter $\Omega_\mathrm{tot}\simeq1$ for the
present universe, the cutoff may be time-independent in the
approximation of the leading.
More specifically, it has been widely accepted that recent observations show that
$\Omega_\mathrm{tot}\simeq1$, $\Omega_m\simeq0.3$, and
$\Omega_\Lambda\simeq0.7$ for the present universe, which
means $|\Omega_k|\ll1$, 
that is, $|\rho_{k0}|\ll\rho_{m0},\rho_{\Lambda0}$, where the
subscript 0 represents the 
values estimated for the present universe scale $a_0$. Moreover, as
shown 
in table~\ref{tab:energy}, $|\rho_k|\sim{a}^{-2}$,
$\rho_m\sim{a}^{-3}$, 
and $\rho_\Lambda\sim{a}^0$. Then, for the past universe, 
i.e. $a<a_0$,
$|\rho_k|=|\rho_{k0}|(a_0/a)^2\ll\rho_{m0}(a_0/a)^3=\rho_m$, and for
the 
future universe, 
i.e. $a>a_0$,
$|\rho_k|=|\rho_{k0}|(a_0/a)^2\ll\rho_{\Lambda0}(a_0/a)^0=\rho_\Lambda$. 
Thus, we get $|\rho_k|\ll\rho_m+\rho_\Lambda\lesssim\rho_\mathrm{tot}$
for 
the whole history of the universe; in other words,
$|\Omega_k|\ll\Omega_\mathrm{tot}$, 
which yields $\Omega_\mathrm{tot}\simeq1$ and $|\Omega_k|\ll1$ for all
eras of the universe,
since $\Omega_\mathrm{tot}+\Omega_k=1$. The density parameters with
respect to the scale 
factor in logarithmic scale are simply  plotted in Fig.~\ref{fig:DP}.
As a result, the cutoff~(\ref{cutoff}) can time-independent in the leading order,
\begin{equation}
\bar{h} \simeq \frac{\ell_P}{\sqrt{90\pi}}. \label{h}
\end{equation}


It is interesting to note 
that there has been a conjecture that the entropy of a system
within a certain boundary is less than or equal to the gravitational
entropy, $S\le S_g=\mathcal{A}/4G$, which is
referred to as the CEB~\cite{bousso}, and the entropy~(\ref{S})
with the cutoff~(\ref{cutoff}) meets the CEB, since the factor~(\ref{F})
is in the range of $0\le\mathcal{F}\le1$ due to the fact that the
equation-of-state parameter~(\ref{eos}) is restricted to
$-1\le\gamma\le1/3$.
If $\gamma<-1$, then
the entropy~(\ref{S}) seems to be incompatible with the CEB because of $\mathcal{F}>1$.
This is because the CEB is valid only when the source energy satisfies
the dominant energy condition
which is violated for the energy of $\gamma<-1$.
On the other hand, our result for the closed universe($k=1$) satisfies
the CEB, which seems to be different from the previous works that
the entropy is not bounded 
for the closed universe \cite{br}. 
The essential reason why the present entropy (\ref{S}) can be bounded even for $k=1$
is that we have considered only the model describing the late-time
accelerated expansion without the big crunch which plays a key role to the unboundedness
of the entropy in the closed universe.

%

In what follows, we would like to investigate the statistical entropy when each
energy is dominant. 
First of all, for the radiation-dominated universe,
$
\Omega_\mathrm{rad}\gg\Omega_m+\Omega_\Lambda$,
that is $\gamma\simeq1/3$, the temperature~(\ref{T}) is very low and
$\mathcal{F}\ll1$ so that the internal energy~(\ref{U}) and the
entropy~(\ref{S}) are very small.
However, it is the early stage of the universe so that
the scale factor might not be sufficiently large in 
applying the present WKB approximation.
Moreover, the behavior of the vanishing entropy looks similar to
that of the extremal black hole so that we have to use other methods to
get desired results.
Second, for the matter-dominated universe,
$
\Omega_m\gg\Omega_\mathrm{rad}+\Omega_\Lambda$
and $|\gamma|\ll1$, the entropy~(\ref{S}) is written as
\begin{equation}
S \simeq \frac{\mathcal{A}}{4^4G},
\end{equation}
and the internal energy~(\ref{U}) is given by $U\simeq3\mathcal{M}/4^5$.
Finally, we consider the vacuum energy-dominated universe,
$
\Omega_\Lambda\gg\Omega_\mathrm{rad}+\Omega_m$
and $\gamma\simeq-1$, then the entropy~(\ref{S}) is rewritten as
\begin{equation}
S \simeq \frac{\mathcal{A}}{4G},
\end{equation}
and the internal energy~(\ref{U}) is given by $U\simeq3\mathcal{M}/4$.
In general, considering power-law cosmology, $\rho_\mathrm{tot}\sim
a^{-n}$ and $\gamma\simeq n/3-1$, the entropy~(\ref{S}) can be written as
\begin{equation}
S = 
\mathcal{F} \frac{\mathcal{A}}{4G},
\end{equation}
and the internal energy~(\ref{U}) is given by
$U\simeq3\mathcal{F}^{4/3}\mathcal{M}/4$, where
$\mathcal{F}=|1-n/4|^3$.
Note that the power-law inflation model, $n\ll1$ leads to the same
entropy with that of the vacuum energy-dominated universe.


In conclusion, motivated by the holographic principle on the
cosmology~\cite{fs}, 
we have studied the entropy of the FRW cosmology 
in terms of the brick
wall method.
As a result, we have shown that 
the entropy is proportional to the area of the apparent horizon and it obeys
the modified area law, $S=\mathcal{FA}/4G$, and 
satisfies the CEB, $S\le\mathcal{A}/4G$, as long as we set the cutoff
as Eq.~(\ref{h}).
Especially, it is interesting to note that 
the maximally saturated entropy satisfying the exact
one-quarter of the area appears at the vacuum
energy-dominated era, which is related to the most promising candidate of the
recently suggested cosmological model to describe the second
accelerated expansion of the universe.

%
The final comment to be mentioned is that 
the Misner-Sharp energy can be
written as $\mathcal{M}=R_A/2G=\rho_\mathrm{tot}\times V_0$, where
$V_0=4\pi R_A^3/3$ is the spatially flat($k=0$) volume. The volume factor
$V_0$ in the Misner-Sharp energy does not change for the $k=\pm1$ case,
which seems somewhat awkward because the volume for the case of $k=\pm1$
is given by $V_\pm\simeq\pi R_A^3$ in the limit of $|\Omega_k|\ll1$, and the
Misner-Sharp energy can not be written in the form of $\mathcal{M}=\rho_\mathrm{tot}\times
V_\pm$ for the $k=\pm1$ case.
However, as commented in Ref.~\cite{br}, the
difference is due to the contribution of the spatial
curvature to the energy. In addition, if we consider the real matter contribution 
as $M=\rho_\mathrm{tot}\times V_\pm$, then the internal
energy~(\ref{U}) becomes $U=M$ for $\mathcal{F}=1$ in the case of $k=\pm1$ and
$|\Omega_k|\ll1$.

\begin{acknowledgments}
W. Kim and E.J. Son were supported by the Korea Science and Engineering Foundation 
(KOSEF) grant funded by the Korea government(MOST) 
(R01-2007-000-20062-0), and M. Yoon was 
supported by the Science Research Center Program of 
the Korea Science and Engineering Foundation 
through the Center for Quantum Spacetime (CQUeST) 
of Sogang University with grant number R11-2005-021.
\end{acknowledgments}



\begin{thebibliography}{99}

\bibitem{thooft:holography}
  G.~'t Hooft,
  \textit{Dimensional reduction in quantum gravity},
  gr-qc/9310026.

\bibitem{bekenstein}
  J.~D.~Bekenstein,
  Lett.\ Nuovo Cim.\  {\bf 4}, 737 (1972);
  Phys.\ Rev.\  D {\bf 7}, 2333 (1973);
  Phys.\ Rev.\  D {\bf 9}, 3292 (1974).

\bibitem{hawking}
  S.~W.~Hawking,
  Commun.\ Math.\ Phys.\  {\bf 43}, 199 (1975)
  [Erratum-ibid.\  {\bf 46}, 206 (1976)].

\bibitem{fs}
  W.~Fischler and L.~Susskind,
  \textit{Holography and cosmology},
  hep-th/9806039.

\bibitem{br}
  D.~Bak and S.-J.~Rey,
  Class.\ Quant.\ Grav.\  {\bf 17}, L83 (2000)
  [hep-th/9902173].

\bibitem{ck}
  R.~G.~Cai and S.~P.~Kim,
  JHEP {\bf 0502}, 050 (2005)
  [hep-th/0501055];
%
  R.~G.~Cai and L.~M.~Cao,
  Phys.\ Rev.\  D {\bf 75}, 064008 (2007)
  [gr-qc/0611071];
%
  M.~Akbar and R.~G.~Cai,
  Phys.\ Rev.\  D {\bf 75}, 084003 (2007)
  [hep-th/0609128].

\bibitem{swc}
  A.~Sheykhi, B.~Wang and R.~G.~Cai,
  Nucl.\ Phys.\  B {\bf 779}, 1 (2007)
  [hep-th/0701198];

  R.~G.~Cai and L.~M.~Cao,
  Nucl.\ Phys.\  B {\bf 785}, 135 (2007)
  [hep-th/0612144].

\bibitem{zrm}
  T.~Zhu, J.~R.~Ren and S.~F.~Mo,
  \textit{Thermodynamics of Friedmann Equation and Masslike Function in Generalized
  Braneworlds},
  arXiv:0805.1162. 

\bibitem{thooft}
  G.~'t Hooft,
  Nucl.\ Phys.\  B {\bf 256}, 727 (1985).

\bibitem{mukohyama} S. Mukohyama, Phys. Rev. D \textbf{61}, 124021 (2000).
\bibitem{su} L. Susskind and J. Uglum, Phys. Rev. D \textbf{50}, 2700
  (1994); T. Jacobson, Phys. Rev. D \textbf{50}, R6031 (1994);
  \textit{Black Hole Entropy and Induced Gravity}, gr-qc/9404039.
\bibitem{dlm} J-G. Demers, R. Lafrance, and R. C. Myers, Phys. Rev. D
  \textbf{52}, 2245 (1995).
\bibitem{gm} A. Ghosh and P. Mitra, Phys. Rev. Lett. \textbf{73}, 2521
  (1994).
\bibitem{kkps} S. W. Kim, W. T. Kim, Y. J. Park, and H. Shin,
  Phys. Lett. B \textbf{392}, 311 (1997).
\bibitem{hkps} J. Ho, W. T. Kim, Y. J. Park, and H. Shin,
  Class. Quantum Grav. \textbf{14}, 2617 (1997).
\bibitem{hzk} F. He, Z. Zhao, and S-W. Kim, Phys. Rev. D \textbf{64},
  044025 (2001); W-B. Liu and Z. Zhao, Chin. Phys. Lett.
  \textbf{18}, 310 (2001). 
\bibitem{zl} Z. Zhou and W. Liu, Int. J. Mod. Phys. A \textbf{19},
  3005 (2004). 

\bibitem{xz}
  L.~Xiang and Z.~Zheng,
  Phys.\ Rev.\  D {\bf 62}, 104001 (2000).

\bibitem{bousso}
  R.~Bousso,
  JHEP {\bf 9907}, 004 (1999)
  [hep-th/9905177].

\bibitem{kreuzer}
  H.~J.~Kreuzer,
  \textit{Non-Equilibrium Thermodynamics and its Statistical Foundations}
  Clarendon Press (1981) 438 p.

\bibitem{carroll}
  S.~M.~Carroll,
  \textit{Spacetime and geometry: An introduction to general relativity},
  San Francisco, USA: Addison-Wesley (2004) 513 p.

\bibitem{hayward}
  S.~A.~Hayward,
  Class.\ Quant.\ Grav.\  {\bf 15}, 3147 (1998)
  [gr-qc/9710089];
%
  S.~A.~Hayward, S.~Mukohyama and M.~C.~Ashworth,
  Phys.\ Lett.\  A {\bf 256}, 347 (1999)
  [gr-qc/9810006].

\bibitem{ms}
  C.~W.~Misner and D.~H.~Sharp,
  Phys.\ Rev.\  {\bf 136}, B571 (1964).

\bibitem{wmap5} For the recent observational data, we refer to the
  five years of WMAP, whose web address is
  \url{http://lambda.gsfc.nasa.gov/product/map/dr3/parameters_summary.cfm}

\end{thebibliography}

\end{document}